\documentstyle[11pt]{article}
\def\bkR{{\rm I\kern-.17em R}}
\def\bkC{{\rm \kern.24em \vrule width.05em height1.4ex depth-.05ex \kern-.26em C}}
\def\bkN{{\rm \kern.50em \vrule width.05em height1.4ex depth-.05ex \kern-.26em N}}

\begin{document}

\author{Jo\~{a}o Nuno Prata\footnote{{Email: joao.prata@mail.telepac.pt}} \\Nuno Costa Dias\footnote{{Email: ncdias@mail.telepac.pt}} \\ {\it Departamento de Matem\'atica} \\
{\it Universidade Lus\'ofona de Humanidades e Tecnologias} \\ {\it Av. Campo Grande, 376, 1749-024 Lisboa, Portugal}\\
{\it and}\\
{\it Grupo de F\'{\i}sica Matem\'atica}\\
{\it Universidade de Lisboa}\\
{\it Av. Prof. Gama Pinto 2}\\
{\it 1649-003 Lisboa, Potugal}}

\title{ENVIRONMENT-INDUCED DECOHERENCE IN NONCOMMUTATIVE QUANTUM MECHANICS\footnote{Presented by J.N.Prata at the {\it Workshop on Advances in Foundations of Quantum Mechanics and Quantum Information with atoms and photons}, 2-5 May 2006, Turin, Italy.}}

\maketitle

\begin{abstract}
We address the question of the appearence of ordinary quantum mechanics in the context of noncommutative quantum mechanics. We obtain the noncommutative extension of the Hu-Paz-Zhang master equation for a Brownian particle linearly coupled to a bath of harmonic oscillators. We consider the particular case of an Ohmic regime. 
\end{abstract}


\section{Introduction}

The recent revival of interest in noncommutative space-time was mainly motivated by the realization that the low energy effective theory of a D-brane in the background of a Neveu-Schwarz-Neveu-Schwarz B field lives on a space with spacial noncommutativity \cite{Seiberg}. A lot of work has since been devoted to investigating quantum field theories on such quantized space-times \cite{Nekrasov}. The most remarkable features of these theories are the breakdown of Poincar\'e invariance, which is replaced by a twisted Poincar\'e symmetry \cite{Wess1} and the so-called IR/UV mixing \cite{Minwalla} in perturbation theory. In most approaches the space-time coordinates $x^{\mu}$ do not commute, either in a canonical way\footnote{We use greek letters $\mu, \nu, \beta, \cdots$ for the space-time indices $0,1,\cdots, d$ and latin letters $i,j,k, \cdots$ for the spatial indices $1, \cdots, d$.}
\begin{equation}
\left[x^{\mu}, x^{\nu} \right] = i \theta^{\mu \nu},\label{1}
\end{equation}
or in a Lie-algebraic way $ \left[x^{\mu}, x^{\nu} \right] = i C_{\beta}^{\mu \nu} x^{\beta}$, where $\theta^{\mu \nu}$ and $C_{\beta}^{\mu \nu}$ are real constants. In general, one assumes that $\theta^{0i}=0$ (or $C_{\beta}^{0i}=0$) (i.e. that time is an ordinary commutative parameter) to avoid problems with the lack of unitarity. Noncommutative quantum mechanics emerges from noncommutative quantum field theory with canonical noncommutativity (\ref{1}) as the non-relativistic one particle sector of the theory. The noncommutative algebra then becomes:
\begin{equation}
\left[\hat q_i, \hat q_j \right] = i \theta_{ij} , \hspace{0.5 cm} \left[\hat q_i, \hat p_j \right] = i \hbar \delta_{ij} , \hspace{0.5 cm} \left[\hat p_i, \hat p_j \right] = 0 ,\label{2} 
\end{equation}
for the spatial indices $i,j=1, \cdots, d$ and time is just a commutative parameter in the theory. The momentum sector is sometimes replaced by the noncommutative generalization \cite{Bertolami}: $\left[\hat p_i, \hat p_j \right]= i \eta_{ij}$, to cast the momenta on equal footing with the coordinates. We shall however consider the simpler version in (\ref{2}) to avert unnecessary technical complications. The algebra (\ref{2}) may also appear in the context of constrained quantum mechanical systems \cite{Girotti}. In our work we shall address the two-dimensional plane with spatial noncommutativity. The algebra (\ref{2}) then reads:
\begin{equation}
\left[\hat q_i, \hat q_j \right] = i \theta \epsilon_{ij} , \hspace{0.5 cm} \left[\hat q_i, \hat p_j \right] = i \hbar \delta_{ij} , \hspace{0.5 cm} \left[\hat p_i, \hat p_j \right] = 0 ,  \hspace{0.5 cm} i,j= 1,2, \label{3}
\end{equation}
where $\epsilon_{12} = - \epsilon_{21} =1$, $\epsilon_{11}= \epsilon_{22}=0$ and $\theta$ is a real constant, henceforth denoted as the noncommutativity parameter. In ref.\cite{Carroll} the authors obtained the following estimate: $\theta \leq 4 \times 10^{-40} m^2$. 

\noindent
To derive the noncommutative version of a quantum mechanical system one first performs a linear transformation (sometimes called the "Seiberg-Witten map"):
\begin{equation}
\hat R_i = \hat q_i + \frac{\theta}{2 \hbar} \epsilon_{ij} \hat p_j, \hspace{0.5 cm} \hat \Pi_i = \hat p_i, \label{4}
\end{equation}
where we adopted the Einstein convention and the set $(\hat R, \hat \Pi)$ satisfy the ordinary  Heisenberg algebra:
\begin{equation}
\left[ \hat R_i , \hat R_j \right] = \left[ \hat \Pi_i , \hat \Pi_j \right]=0 , \hspace{0.5 cm} \left[ \hat R_i , \hat \Pi_j \right] = i \hbar \delta_{ij} . \label{5}
\end{equation}
The transformation (\ref{4}) is obviously not unique, as the canonical set $(\hat R, \hat \Pi)$ are always defined only up to a unitary transformation. Once the Hamiltonian $\hat H (\hat q, \hat p)$ has been expressed in terms of the canonical variables $(\hat R, \hat \Pi)$, the usual quantization procedure follows. In particular, the space of states of noncommutative quantum mechanics remains $L^2 \left( \bkR^2, dR \right)$.

Our aim is to study the emergence of ordinary quantum mechanics in the context of noncommutative quantum mechanics. The question of this noncommutative-commutative transition (NC-C) is as delicate as that of the quantum-classical transition \cite{Zeh}. In fact we should expect various criteria, such as formal limits $\theta \to 0$, high quantum numbers $n \to 0$, certain high-temperature expansions or resorting to suitable coherent states, etc. In many cases one may even be forced to combine various of these criteria.

Our strategy to induce a NC-C transition consists of coupling a noncommutative Brownian oscillator to an external reservoir of noncommutative oscillators at thermal equilibrium and thus treat it as an open system \cite{Hu3}. We shall assume the coupling to be weak so that only the linear response will be considered \cite{Caldeira}. The Hamiltonian reads:
\begin{equation}
\hat H = \frac{\hat p^2}{2 M} + \frac{1}{2} M \Omega^2 \hat q^2 + \sum_n \left( \frac{ \left(\hat p^{(n)} \right)^2}{2 m_n} + \frac{1}{2} m_n \omega_n^2 \left(\hat q^{(n)} \right)^2 \right) + \sum_n C_n \hat q \cdot \hat q^{(n)}, \label{6}
\end{equation}
where $\hat q= (\hat q_1 , \hat q_2)$, $\hat p= (\hat p_1 , \hat p_2)$ are the position and momentum of the Brownian oscillator of mass $M$ and bare frequency $\Omega$, $\hat q^{(n)}= (\hat q_1^{(n)} , \hat q_2^{(n)})$, $\hat p^{(n)}= (\hat p_1^{(n)} , \hat p_2^{(n)})$ are the positions and momenta of the bath's oscillators with masses $m_n$ and frequencies $\omega_n$. We have also used the notation $\hat q^2 = \hat q_1^2 + \hat q_2^2$, $\hat q \cdot \hat q^{(n)} = \hat q_1 \hat q_1^{(n)} + \hat q_2 \hat q_2^{(n)}$, etc. 

We shall work in quantum phase space by resorting to the so-called quasi-distribution (or deformation quantization) formulation of quantum mechanics. Obviously we shall have to implement some extension of this formulation to incorporate the noncommutativity. For instance the noncommutative $\star$-product is replaced by the so-called super $\star$-product \cite{Hu3}:
\begin{equation}
A(z) \star B (z) = \left. \exp \left( \frac{i \hbar}{2} \nabla_z  \cdot J \nabla_{z'} + \frac{i \theta}{2} \nabla_q \cdot \epsilon \nabla_{q'} \right) A (z) B (z') \right|_{z'=z}. \label{7}   
\end{equation}
with $\nabla_q = \left( \partial / \partial q_1, \partial / \partial q_2 \right)$ and $\epsilon_{ij}$ is as in eq.(3). Moreover, $\nabla_z = \left( \nabla_q,  \nabla_p \right)$, $z= (q,p)$, etc and $J$ is the $4 \times 4$ symplectic matrix:
\begin{equation}
J = \left(
\begin{array}{c r}
0_{2 \times 2} & -1_{2 \times 2}\\
1_{2 \times 2} & 0_{2 \times 2}
\end{array}
\right) \label{8}
\end{equation}
The Wigner function also has to be modified. For a pure state it reads \cite{Jing}:
\begin{equation}
F_{pure}^{NC} (q,p) = \frac{1}{(\pi \hbar)^2} \int dy \hspace{0.2 cm} e^{-2 i p \cdot y / \hbar} \psi (q+y) \star_{\theta} \psi^* (q-y), \label{9}
\end{equation}
where $\psi$ is the wave-function and $\star_{\theta}$ is the noncommutative $\star$-product which is obtained from Eq.~(\ref{7}) by setting $\hbar =0$. The important thing to remark is that this $\star_{\theta}$-product only involves the coordinates $q$.

\noindent
A mixed state is just a convex combination of pure states. The dynamics of the noncommutative Wigner function is governed by the super Moyal equation:
\begin{equation}
\frac{\partial F^{NC}}{\partial t}(z,t) = \left[H (z) , F^{NC} (z,t) \right]. \label{10}
\end{equation}
The super-Moyal bracket is defined by: $i \hbar \left[A(z) , B (z) \right] = A (z) \star B (z) - B(z) \star A (z)$. 
In Eqs.~(\ref{7}),~(\ref{10}) $z$ may now represent an array of positions and momenta for several particles. For our system, Eq.~(\ref{10}) describes the dynamics of the full combined (closed) system of the Brownian particle and the reservoir. We are interested in the master equation for the reduced noncommutative Wigner function of the Brownian particle:
\begin{equation}
W(z)  \equiv \int \left( \Pi_n dz^{(n)} \right) F^{NC} \left(z, \left\{z^{(n)} \right\} \right), \label{11}
\end{equation}
where $dz^{(n)} = dq^{(n)} d p^{(n)}$. To simplify the derivation we assume that the initial distributions of the Brownian particle and of the bath are uncorrelated:
\begin{equation}
F^{NC} \left(z, \left\{z^{(n)} \right\} , t=0 \right)  =W (z, t=0) W^b \left( \left\{z^{(n)} \right\} , t=0 \right), \label{12}
\end{equation}
and that the bath is at thermal equilibrium:
\begin{equation}
W^b \left(\left\{  z^{(n)}  \right\} , t=0  \right) = \prod_n N_n 
\times \exp \left[- a_n \left(p^{(n)} \right)^2 - c_n  \left(q^{(n)} \right)^2 - 2 b_n  L^{(n)} \right], \label{13}
\end{equation}
where $L^{(n)} =  q^{(n)} \cdot \epsilon p^{(n)}$ and $a_n, b_n, c_n, N_n$ are certain temperature dependent, positive coefficients. After a lengthy calculation which follows closely the method of Halliwell and Yu \cite{Halliwell} we arrive \cite{Dias5} at the noncommutative extension of the Hu-Paz-Zhang master equation \cite{Hu1}: 
\begin{equation}
\begin{array}{c}
\frac{\partial W}{\partial t} = - \frac{p}{M} \cdot \nabla_q W + M \Omega^2 q \cdot \nabla_p W + \left( \nabla_p W \right) \cdot A(t) q + \nabla_p  \cdot \left( B(t) p W \right) + \\
\\
+ \nabla_p  \cdot \left( C(t) \nabla_q W \right) + \nabla_p  \cdot \left( D(t) \nabla_p W \right) + \frac{\theta}{\hbar}  M \Omega^2 q  \cdot \epsilon  \nabla_q W -  \frac{\theta}{\hbar}  \nabla_q \cdot \epsilon \left(A(t) q W \right) \\
\\
-  \frac{\theta}{\hbar}  \nabla_q \cdot \epsilon \left(B(t) p W \right) -  \frac{\theta}{\hbar}  \nabla_q \cdot \epsilon \left(C(t) \nabla_q W \right) -  \frac{\theta}{\hbar}  \nabla_q \cdot \epsilon \left(D(t) \nabla_p W_r \right),
\end{array} \label{14}
\end{equation}
where $A, B, C, D$ are time dependent $2 \times 2 $ matrices. Their explicit form depends solely on the so-called dissipation and noise kernels. In the Heisenberg picture the equation of motion for the Brownian particle is the noncommutative Langevin equation:
\begin{equation}
\begin{array}{c}
\ddot Q_i (t) + \Omega^2 Q_i (t) - \frac{\theta}{\hbar} M \Omega^2 \epsilon_{ij} \dot Q_j (t) + \\
\\
+  \frac{2}{M} \int_0^t ds \hspace{0.2 cm} \eta_{kj} (t-s) \left(\delta_{ik} - \frac{M \theta}{\hbar} \epsilon_{ik} \frac{d}{ds} \right) Q_j (s) = \frac{f_i (t)}{M}, 
\end{array}
\label{15}
\end{equation}
where $Q_i  = q_i + \frac{\theta}{\hbar} \epsilon_{ij} p_j$ is used for convenience. The dissipation matrix kernel $\eta_{ij} (t) = \frac{d}{dt} \gamma_{ij} (t)$ is given by:
\begin{equation}
\gamma_{ij} (t) = \int_0^{+ \infty} \frac{d \omega}{\omega} \left[ I_{ij} (\omega) \cos (\omega t) + J_{ij} (\omega) \sin (\omega t) \right]. \label{16}
\end{equation}
The spectral densities read:
\begin{equation}
\left\{
\begin{array}{l}
I_{ij} (\omega) = \sum_n \frac{C_n^2 \omega^2 }{4 m_n \omega_n^2 \Omega_n } \delta_{ij} \left[\delta (\omega - \Omega_n - \lambda_n ) + \delta (\omega - \Omega_n + \lambda_n ) \right]\\
\\
J_{ij} (\omega) = \sum_n \frac{C_n^2 \omega^2 }{4 m_n \omega_n^2 \Omega_n } \epsilon_{ij} \left[\delta (\omega - \Omega_n - \lambda_n ) - \delta (\omega - \Omega_n + \lambda_n ) \right]
\end{array}
\right. \label{17}
\end{equation}
where $\Omega_n = \omega_n \sqrt{1+ ( \lambda_n/ \omega_n)^2}$ and $\lambda_n = m_n \omega_n^2 \theta / (2 \hbar)$. Finally, the "random" force $f_i (t)$ satisfies:
\begin{equation}
<f_i (t)> =0, \hspace{1 cm} < \left\{f_i (t), f_j (t') \right\}>= \hbar \nu_{ij} (t -t'), \label{18}
\end{equation}
where $\left\{A,B \right\} = \left( AB +BA \right)/2$ is the anticommutator and 
\begin{equation}
\nu_{ij} (t) = \int_0^{+ \infty} d \omega \hspace{0.2 cm} \coth \left( \frac{\hbar \beta \omega}{2} \right) \left[I_{ij} (\omega) \cos (\omega t) + J_{ij} (\omega) \sin (\omega t) \right] \label{19}
\end{equation}
is the noise kernel. The two kernels satisfy the fluctuation-dissipation relation:
\begin{equation}
\nu_{ij} (t) = \int_{- \infty}^{+ \infty} ds \hspace{0.2 cm} K(t-s) \gamma_{ij} (s), \hspace{1 cm} K(t) = \int_0^{+ \infty} \frac{d \omega}{\pi} \hspace{0.2 cm} \coth \left( \frac{\hbar \beta \omega}{2} \right) \cos (\omega t). \label{20}
\end{equation} 
Given the kernels $\eta$ and $\nu$ we were able to write an expression for the matrix coefficients $A,B,C,D$ of the master equation (\ref{14}). These expressions simplify drastically if we consider the weak coupling limit and neglect terms of order higher than ${\cal O} \left(C_n^2 \right)$ in the coupling constants. In Ref.\cite{Dias5} we wrote the expression for the coefficients in this weak coupling limit. Here we wish to report on another particular case of interest, namely the noncommutative version of the Caldeira-Leggett model \cite{Caldeira}. In this model, to achieve irreversibility, we shall take the thermodynamic limit, i.e. we consider a continuum of oscillators,
\begin{equation}
\sum_n C_n^2 \longrightarrow \int_0^{+ \infty} d \omega \hspace{0.2 cm} \rho_D (\omega) C^2 (\omega), \label{21}
\end{equation}
with density $\rho_D (\omega)$ such that $\pi \rho_D (\omega) C^2 ( \omega) = \theta (\Lambda- \omega) 2 m \eta \omega^2$, where $\theta (\omega)$ is Heaviside's step function, $\Lambda$ is a high frequency cutoff which shall eventually be taken to infinity, $m$ is a characteristic mass of the bath $(m_n \approx m, \forall n)$ and $\eta$ is a damping constant. This model is called ohmic in the sense that (in the commutative limit) the spectral density reads (cf.Eq.~(\ref{17})): $I_{ij} (\omega) \sim \frac{\eta}{\pi} \delta_{ij} \omega$. In this ohmic regime we get from Eq.~(\ref{16}) to first order in $\theta$:
\begin{equation}
\eta_{ij} (t) \sim \eta \left( \delta_{ij} + \frac{m}{2} \frac{\theta}{\hbar} \epsilon_{ij} \frac{d}{dt} \right) \delta (t), \label{22}
\end{equation}
where we used: $\lim_{\Lambda \to + \infty} \frac{\sin (\Lambda t)}{\pi t} = \delta (t)$. As in \cite{Caldeira} we now assume a high temperature ohmic regime, so that in (\ref{19}) we shall consider $\hbar \Omega << k_B T$. We shall thus only admit the first term in the expansion of the coth in Eq.~(\ref{19}). It then follows that
\begin{equation}
\nu_{ij} (t) \sim \frac{2 \eta k_B T}{\hbar} \left( \delta_{ij} + \frac{m}{2} \frac{\theta}{\hbar} \frac{d}{dt} \right) \delta (t). \label{23}
\end{equation}
We thus get after some algebra:
\begin{equation}
\begin{array}{c}
\frac{\partial W}{\partial t}  \sim - \frac{p_i}{M} \frac{\partial W}{\partial q_i} + M \Omega_{ren}^2 q_i \frac{\partial W}{\partial p_i } - \frac{2 \eta}{M} \frac{\partial}{\partial p_i} \left( p_i W \right) + 2 \eta k_B T \frac{\partial^2 W}{\partial p_i^2}\\
\\
- \frac{\theta}{\hbar} \kappa \epsilon_{ij} q_i \frac{\partial W}{\partial p_j} - \frac{\theta}{\hbar} \xi \epsilon_{ij} \frac{\partial}{\partial p_i} \left(p_j W \right) - \frac{\theta}{\hbar} \zeta k_B T \epsilon_{ij} \frac{\partial^2 W}{\partial q_i \partial p_j} + \\
\\
+ \frac{\theta}{\hbar} M \Omega_{ren}^2 \epsilon_{ij} \frac{\partial}{\partial q_i} \left(q_j W \right) + \frac{2 \eta}{M} \frac{\theta}{\hbar} \epsilon_{ij} \frac{\partial}{\partial q_i}  \left(p_j W \right),
\end{array} \label{24}
\end{equation}
where $\Omega_{ren}^2 = \Omega^2 - \frac{2 \eta}{M} \delta (0)$ is the renormalized frequency and $\kappa$, $\xi$, $\zeta$ are phenomenological constants which depend on the dissipation coefficient $\eta$ and on $m$, $M$ and $\Omega_{ren}^2$.

\vspace{0.3 cm}
\noindent
Equation (\ref{24}) will be the object of futher investigation in a forthcoming paper.
 
\begin{center}

{\large{{\bf Acknowledgments}}} 

\end{center}

\vspace{0.3 cm}
\noindent
The authors wish to thank C. Bastos, O. Bertolami and A. Mikovic for useful comments. This work was partially supported by the grants POCTI/MAT/45306/2002 and POCTI/0208/2003 of the Portuguese Science Foundation.

\end{document}